\documentclass [12pt]{article}
\usepackage{graphicx}

\begin{document}

\textbf{Approximation of scattering phases for Reid93 potential}

\textbf{V. I. Zhaba}

Uzhgorod National University, Department of Theoretical Physics,

54, Voloshyna St., Uzhgorod, UA-88000, Ukraine

\textbf{Abstract}

For a single-channel nucleon-nucleon scattering, a well-known and
convenient variable phase approach is considered, which is widely
used for practical problems of atomic and nuclear physics.
Approximation of the \textit{pp}- and \textit{np}- scattering
phases obtained for the modern realistic phenomenological
nucleon-nucleon potential Reid93 was carried out. The
approximation function is used as a well-known formula for a
parabolic-type quadratic function.


\textbf{1. Introduction}

From the experimentally observed values, first of all, information
is obtained about the phase and amplitude of scattering, rather
than the wave functions. The latter are the main object of
research in the standard approach. That is, in the experiment
there are not the wave functions themselves, but their changes
caused by the interaction \cite{Babikov1988}. Therefore, it is
interesting to obtain and use such an equation that directly
bundles the phase and scattering amplitudes with potential and
does not find the wave functions at the same time.

The exact solution of the scattering problem for the purpose of
calculating the scattering phases is possible only for individual
phenomenological potentials. For realistic potentials, the
scattering phase is approximated. This is due to the use of
physical approximations or numerical calculations. The influence
of the choice of a numerical algorithm on the solution of the
scattering problem is given in \cite{Haysak2009}.

The main methods for solving the Schr\"{o}dinger equation for the purpose of
obtaining scattering phases are: the method of successive approximations,
the Born approximation, the variable phase approach (or phase-function
method) and others.

In previous papers \cite{Zhaba2015, Zhaba20161, Zhaba20162,
Zhaba20163} we obtain scattering phases for a set of
nucleon-nucleon potentials. The variable phase approach (VPA) was
used to find phase shifts. On the received scattering phases were
calculated scalar scattering amplitude, the full cross-section and
the partial scattering amplitude

This paper deals with the analysis of calculated phase shifts of
nucleon-nucleon scattering in various spin states for the modern
realistic phenomenological nucleon-nucleon potential Reid93 by
using the variable phase approach (results according to paper
\cite{Zhaba20164}) and phase shifts from the original paper
\cite{Stoks1994} for the same potential.

\textbf{2. The variable phase approach}

In the scattering of a spin-free particle with energy $E$ and an
orbital moment $l$ on the spherical-symmetric potential $V(r)$,
the Schr\"{o}dinger equation for the radial wave function
$u_{l}(r)$ has the form \cite{Babikov1988}:

\begin{equation}
\label{eq1}
u''_l (r) + \left( {k^2 - \frac{l(l + 1)}{r^2} - U(r)} \right)u_l (r) = 0,
\end{equation}

where $U(r) = 2mV(r) / \hbar ^2$ - the renormalized interaction potential,
$m$ - the reduced mass, $k^2 = 2mE / \hbar ^2$ - the wave number.

VPA is a extra, specific and special method for solving the Schr\"{o}dinger
radial equation (\ref{eq1}). This method is convenient for obtaining scattering
phases. This is due to the fact that this method does not need to first
calculate radial wave functions in a wide area and then, by their asymptotic
behavior, find these phases.

Two linearly independent solutions of the free Schr\"{o}dinger equation (\ref{eq1})
are the known Riccati-Bessel functions $j_{l}$\textit{(kr)} and $n_{l}$\textit{(kr)}. The free motion is
only responsible for the solution $j_{l}$\textit{(kr)} at the point $r$=0. In this case, the
solution becomes asymptotically large for $r$ values

\[
u_l (r) \approx const \cdot \sin (kr - l\pi / 2).
\]

The presence of the potential leads to the fact that now, in the field of
potential disappearance $U(r)$, the wave function includes an additive
$n_{l}$\textit{(kr)} for an irregular solution of the free equation. And in particular,
the measure of this additive is the scattering phase \textit{$\delta $}$_{l}$:

\[
u_l (r) \approx const \cdot \left[ {j_l (kr) - tg\delta _l \cdot n_l (kr)}
\right],
\]

$u_l (r) \to const \cdot \sin (kr - l\pi / 2 + \delta _l ),r \to \infty $.

A standard and generally accepted method for calculating
scattering phases is the direct solution of the Schr\"{o}dinger
equation (\ref{eq1}) with an asymptotic boundary condition. VPA is
the transition from the Schr\"{o}dinger equation (\ref{eq1}) to
the equation for the phase function. To do this, make the
following simple replacement \cite{Babikov1988, Babikov1967}:

\begin{equation}
\label{eq2}
u_l (r) = A_l (r)\left[ {\cos \delta _l (r) \cdot j_l (kr) - \sin \delta _l
(r) \cdot n_l (kr)} \right].
\end{equation}

The two new functions \textit{$\delta $}$_{l}(r)$ and $A_{l}(r)$
introduced have the physical content of the corresponding
scattering phases and the rationing constants (or amplitudes) of
wave functions for scattering on a definite sequence of truncated
potentials. They are called according to their physical content by
phase and amplitude function. The term "phase function" was first
used in the paper of Morse and Allis \cite{Morse1933}. The
equation for phase and amplitude functions with initial conditions
is written in this form \cite{Babikov1988, Babikov1967}:

\begin{equation}
\label{eq3}
\delta '_l = - \frac{1}{k}U\left[ {\cos \delta _l \cdot j_l - \sin \delta _l
\cdot n_l } \right]^2,\mbox{ }\delta _l (0) = 0;
\end{equation}

\begin{equation}
\label{eq4}
A'_l = - \frac{1}{k}A_l U\left[ {\cos \delta _l \cdot j_l - \sin \delta _l
\cdot n_l } \right]\left[ {\sin \delta _l \cdot j_l + \cos \delta _l \cdot
n_l } \right],\mbox{ }A_l (0) = 1.
\end{equation}

The phase equation (\ref{eq3}) was first obtained by Drukarev
\cite{Drukarev1949}, and then independently in the papers of
Bergmann, Kynch \cite{Kynch1952}, Olson, Calogero
\cite{Calogero1963} and Dashen \cite{Dashen1963}. VPA proved to be
convenient in solving many practical problems of atomic and
nuclear physics.

\textbf{3. Approximation of scattering phases}

In the original paper \cite{Stoks1994}, scattering phases were
obtained for the potentials of Nijmegen group (Nijm I, Nijm II and
Reid93) and for Nijmegen multienergy partial-wave analysis. In
\cite{Zhaba20164}, phase shifts \textit{nn}-, \textit{pp}-,
\textit{np}- scattering for the nucleon-nucleon potential Reid93
were obtained using the variable phase approach.

According to a detailed analysis in \cite{Zhaba20164}, we can draw
the following conclusions. Comparison of phase shifts for
\textit{pp}- and \textit{np}- scattering calculated for the same
Reid93 potential by different methods indicates that the
difference between the results is not more than two percent.
Comparison of the results of phase shift calculations for the
Reid93 potential obtained with VPA and phase shifts for other
potential models (NijmI, NijmII \cite{Stoks1994}, Argonne v18
\cite{Wiringa1995} and CD-Bonn \cite{Machleidt2001}) and for
partial wave analysis \cite{Stoks1994} indicates that the
deviation between these data is up to five percent. The results of
calculations of single-channel scattering phases for the Reid93
potential are in good agreement with the data obtained in the
framepaper of the chiral perturbation theory \cite{Epelbaum2015}
and for the partial wave analysis below the pion formation
threshold \cite{Perez2013}.

The structure of the Paris potential in paper
\cite{Dolgopolov2017} is analyzed from the point of view of the
independence of the coefficients of various components. In
addition, in \cite{Dolgopolov2017} the approximation of the
scattering phases obtained for the Paris potential is carried out.

In this paper, the approximation of the phase's of \textit{pp}-
and \textit{np}- scattering from papers \cite{Zhaba20164} and
\cite{Stoks1994} for the Reid93 potential is carried out. A
parabolic-type quadratic function was used, which Dolgopolov,
Minin and Rabotkin used for scattering phases for the Paris
potential \cite{Dolgopolov2017}

\begin{equation}
\label{eq5}
y_i (x) = a + bx + cx^2.
\end{equation}

The obtained coefficients $a$, $b$, $c$ for the approximation of phases \textit{pp}- and \textit{np}-
scattering are given in Tables 1 and 2, respectively.

The following values are calculated for the estimation of the quality of
approximation of the phase scattering $\delta _i $:

1) standard deviation of the fit:

\[
\sigma = \sqrt {\frac{\sum\limits_{i = 1}^N {\left( {\delta _i - y_i (x)}
\right)^2} }{N - P}} ;
\]

2) \textit{$\chi $}$^{2}$ per degree of freedom of function;

3) correlation coefficient $R$.

According to the data of Tables 1 and 2 within a single spin state, it is
difficult to find the difference between the coefficients for the
approximation form (\ref{eq5}) and the approximation parameters. Only one can
estimate for which of the states the approximation with the quadratic
function (\ref{eq5}) will be "better" or "worse". In the end, the minimum value of
the correlation coefficient will be for the $^{3}$P$_{0}$- state (for both
\textit{pp}- scattering and \textit{np}- scattering), and the maximum value - for the
$^{1}$D$_{2}$- state.

\textbf{Table 1}. Parameters of approximation for \textit{pp}-
scattering phases for the Reid93 potential of papers
\cite{Zhaba20164, Stoks1994}

\begin{tabular}{|l|l|l|l|l|l|l|}
\hline State& $\sigma $& $\chi ^{2} $& $R$& $a$& $b$&
$c$ \\
\hline $^{1}$S$_{0}$ \cite{Stoks1994}& 7.23422& 52.33388& 0.96423&
50.21339& -26.42685&
2.51043 \\
\hline $^{1}$S$_{0}$ \cite{Zhaba20164}& 7.10054& 50.41768&
0.96600& 50.54306& -27.22381&
2.71849 \\
\hline $^{3}$P$_{0}$ \cite{Stoks1994}& 3.49740& 12.23184& 0.92675&
4.04607& 6.15318&
-3.52370 \\
\hline $^{3}$P$_{0}$ \cite{Zhaba20164}& 3.51237& 12.33673&
0.92790& 4.10873& 6.08925&
-3.53163 \\
\hline $^{3}$P$_{1}$ \cite{Stoks1994} & 0.75864& 0.57554& 0.99819&
-0.87440& -13.22648&
1.38537 \\
\hline $^{3}$P$_{1}$ \cite{Zhaba20164} & 0.76060& 0.57851&
0.99822& -0.91270& -13.43629&
1.42428 \\
\hline $^{1}$D$_{2}$ \cite{Stoks1994}& 0.12204& 0.01489& 0.99963&
-0.24939& 4.41924&
-0.39851 \\
\hline $^{1}$D$_{2}$ \cite{Zhaba20164}& 0.12482& 0.01558& 0.99962&
-0.25045& 4.49823&
-0.41495 \\
\hline
\end{tabular}

\textbf{Table 2}. Parameters of approximation for \textit{np}-
scattering phases for the Reid93 potential of papers
\cite{Zhaba20164, Stoks1994}

\begin{tabular}{|l|l|l|l|l|l|l|}
\hline State& $\sigma $& $\chi ^{2} $& $R $& $a$& $b$&
$c$ \\
\hline $^{1}$S$_{0}$ \cite{Stoks1994}& 2.53466& 6.42449& 0.99674&
61.73167& -38.2194&
5.19865 \\
\hline $^{1}$S$_{0}$ \cite{Zhaba20164}& 2.52679& 6.38469& 0.99677&
61.74440& -38.22495&
5.19380 \\
\hline $^{3}$P$_{0}$ \cite{Stoks1994}& 3.34895& 11.21545& 0.93174&
3.96439& 5.64195&
-3.37986 \\
\hline $^{3}$P$_{0}$ \cite{Zhaba20164}& 3.35480& 11.25466&
0.93191& 3.97839& 5.63373&
-3.38344 \\
\hline $^{1}$P$_{1}$ \cite{Stoks1994}& 1.01931& 1.03899& 0.99625&
-1.53013& -14.91513&
2.14592 \\
\hline $^{1}$P$_{1}$ \cite{Zhaba20164}& 1.02139& 1.04324& 0.99624&
-1.53073& -14.95112&
2.15576 \\
\hline $^{3}$P$_{1}$ \cite{Stoks1994}& 0.71160& 0.50637& 0.99846&
-0.85135& -13.34335&
1.37341 \\
\hline $^{3}$P$_{1}$ \cite{Zhaba20164}& 0.71104& 0.50557& 0.99847&
-0.85390& -13.36223&
1.37541 \\
\hline $^{1}$D$_{2}$ \cite{Stoks1994}& 0.13709& 0.01879& 0.99956&
-0.26844& 4.52958&
-0.40765 \\
\hline $^{1}$D$_{2}$ \cite{Zhaba20164}& 0.13693& 0.01875& 0.99956&
-0.27077& 4.55897&
-0.41614 \\
\hline $^{3}$D$_{2}$ \cite{Stoks1994}& 0.91072& 0.82941& 0.99723&
-0.48671& 20.00056&
-3.76889 \\
\hline $^{3}$D$_{2}$ \cite{Zhaba20164}& 0.91915& 0.84484& 0.99718&
-0.47793& 20.02265&
-3.77559 \\
\hline
\end{tabular}

On Fig. 1-3 shows the phases of \textit{np}- scattering from paper
\cite{Zhaba20164} (points) and the results of approximation
(curve). The illustration only shows data for three states
($^{1}$S$_{0}$, $^{3}$P$_{1}$, $^{3}$D$_{2})$.

\pdfximage width 111mm {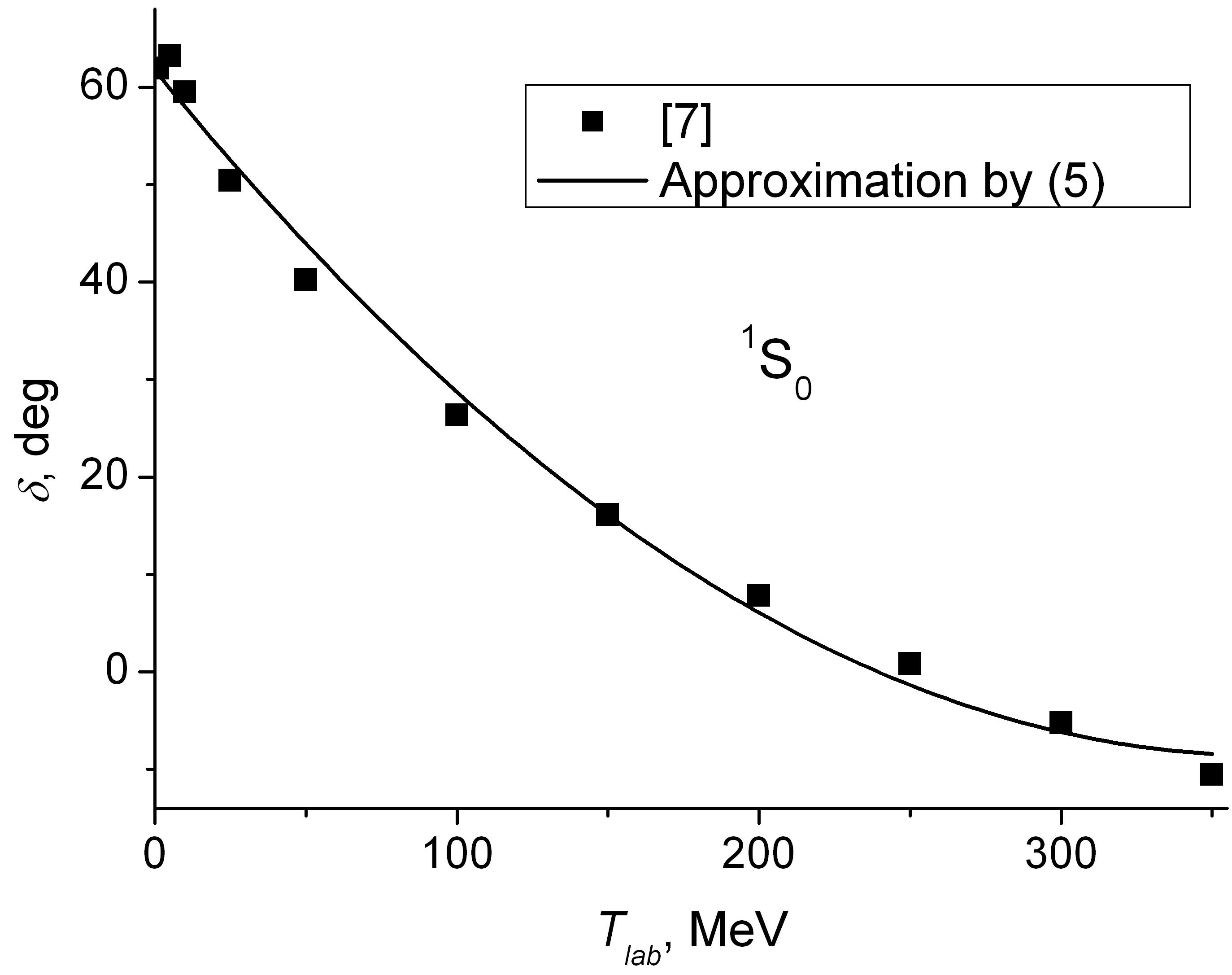}\pdfrefximage\pdflastximage

Fig.~1. Phases of \textit{np}- scattering for $^{1}$S$_{0}$- state

\pdfximage width 111mm {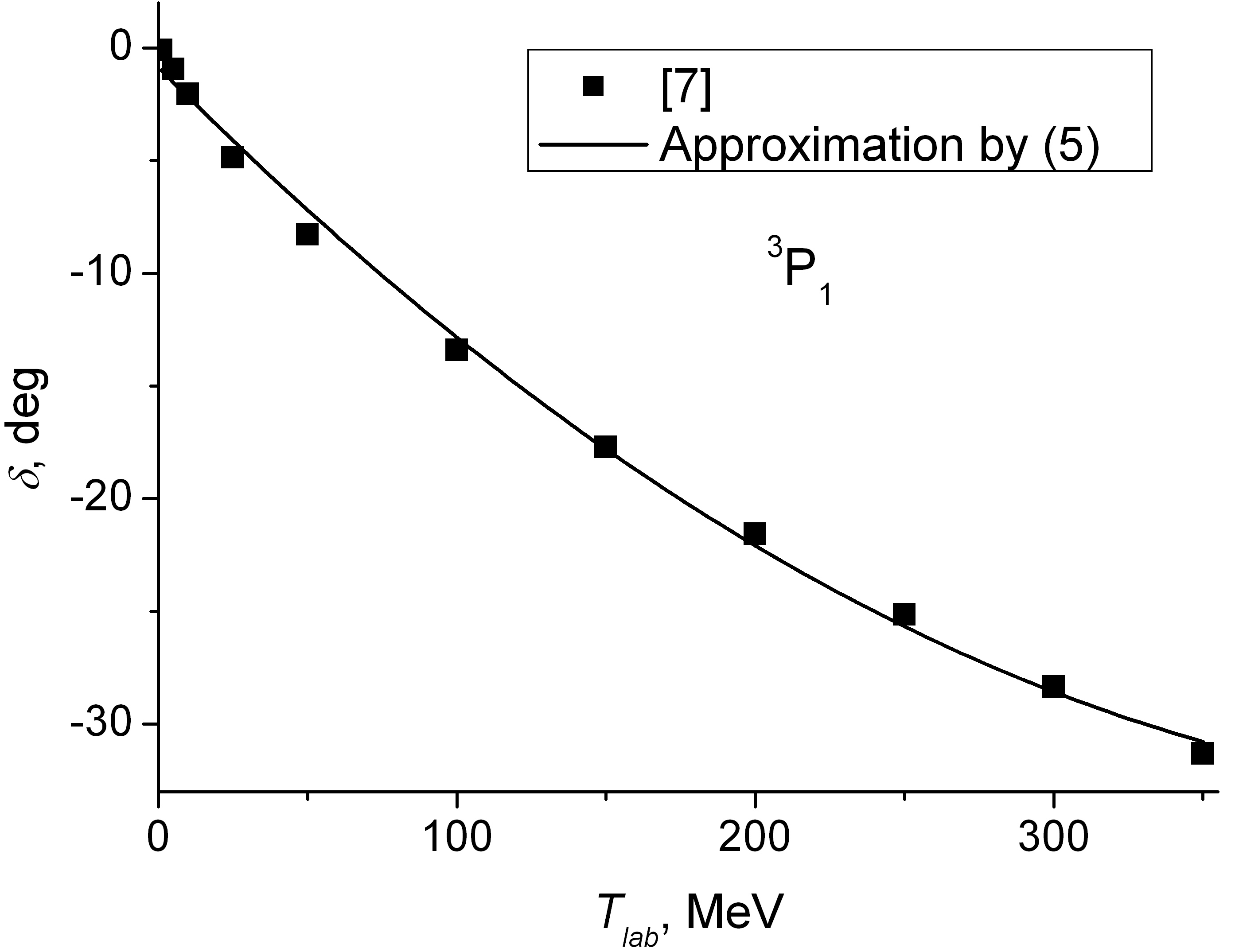}\pdfrefximage\pdflastximage

Fig.~2. Phases of \textit{np}- scattering for $^{3}$P$_{1}$- state

\pdfximage width 111mm {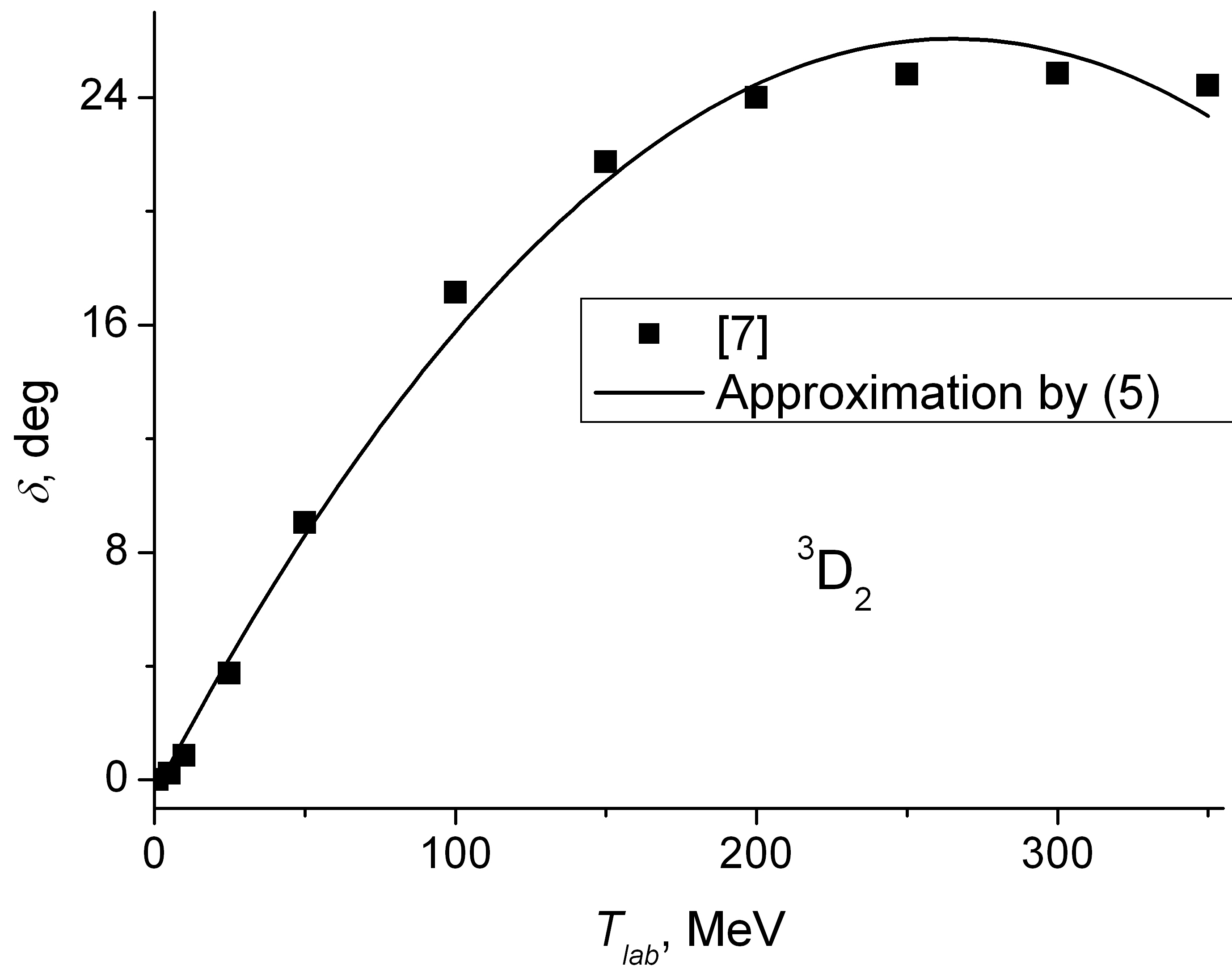}\pdfrefximage\pdflastximage

Fig.~3. Phases of \textit{np}- scattering for $^{3}$D$_{2}$- state

\textbf{4. Conclusions}

The paper considers the variable phase approach for the problem of
single-channel nucleon-nucleon scattering and the final results of the
application of this method for the search of scattering phases for a
specific Reid93 interaction potential.

For the approximation of the scattering phases, we use a
parabolic-type quadratic function in the form (\ref{eq5}), which
was proposed in paper \cite{Dolgopolov2017}. The results of the
approximation of the \textit{pp}- and \textit{np}- scattering
phases obtained by different methods are compared.

The form of the recording of the phase function in a convenient
form (\ref{eq5}) allows it to be used for further calculation or
recalculation of the scalar scattering amplitude, the full
cross-section and the partial scattering amplitude accordingly
\cite{Babikov1988}

\begin{equation}
\label{eq6}
F(\theta ) = \frac{1}{k}\sum\limits_{l = 0}^\infty {(2l + 1)e^{i\delta _l
}\sin \delta _l P_l (\cos \theta )} ,
\end{equation}

\begin{equation}
\label{eq7}
\sigma = \frac{4\pi }{k^2}\sum\limits_{l = 0}^\infty {(2l + 1)\sin ^2\delta
_l } ,
\end{equation}

\begin{equation}
\label{eq8}
f_l = \frac{1}{k}e^{i\delta _l }\sin \delta _l ,
\end{equation}

For calculations, it will not be necessary to calculate the scattering
phases separately for each value. And, having a specific approximate
function for the phase function, it will be possible to calculate the values
of (\ref{eq6})-(\ref{eq8}) for any phase value within the limits of the approximation
carried out (in the interval $T_{lab}$=0-350 MeV).

In further researches one can obtain and compare coefficients and parameters
for approximation for other modern phenomenological potentials.


\begin{thebibliography}{0}

\bibitem{Babikov1988} V.V. Babikov, \textit{The phase-function method in
quantum mechanics} (Moscow, Science, 1988).

\bibitem{Haysak2009} I. Haysak and V. Zhaba, Visnyk Lviv Univ. Ser. Phys. \textbf{44}, 8
(2009).

\bibitem{Zhaba2015} V.I. Zhaba, Calculation of phases of nucleon-nucleon scattering
for potentials NijmI, NijmII and Nijm93 on the phase-function
method, J. Phys. Stud. \textbf{19}, 4201 (2015).

\bibitem{Zhaba20161} V.I. Zhaba, The phase-functions method and full cross-section
of nucleon-nucleon scattering, Mod. Phys. Lett. A \textbf{31},
1650049 (2016).

\bibitem{Zhaba20162} V.I. Zhaba, The phase-functions method and scalar amplitude of
nucleon-nucleon scattering, International Journal of Modern
Physics E \textbf{25}, 1650088 (2016).

\bibitem{Zhaba20163} V.I. Zhaba, Calculation of phases of np-scattering up to Tlab=3
GeV for Reid68 and Reid93 potentials on the phase-function method,
Probl. Atom. Sci Tech. \textbf{5(105)}, 29 (2016).

\bibitem{Zhaba20164} V. Zhaba, Calculation of phases of nucleon-nucleon scattering
for potential Reid93 on the phase-functions method, Scientific
notes NaUKMA, Physics and mathematics \textbf{178}, 44 (2016).

\bibitem{Stoks1994} V.G.J. Stoks, R.A.M. Klomp, C.P.F. Terheggen, J.J. de Swart,
Construction of high quality NN potential models, Phys. Rev. C
\textbf{49}, 2950 (1994).

\bibitem{Babikov1967} V.V. Babikov, The phase-function method in quantum mechanics,
Sov. Phys. Usp. \textbf{10}, 271 (1967).

\bibitem{Morse1933} P.M. Morse and W. P. Allis, The Effect of Exchange on the
Scattering of Slow Electrons from Atoms, Phys. Rev. \textbf{44},
269 (1933).

\bibitem{Drukarev1949} G.F. Drukarev, About determination of phase of wave function
at dispersion of particles, ZhETF \textbf{19}, 247 (1949).

\bibitem{Kynch1952} G.I. Kynch, The Two-Body Scattering Problem with Non-Central
Forces I - Non-Relativistic, Proc. Phys. Soc. A \textbf{65}, 83
(1952).

\bibitem{Calogero1963} F. Calogero, A novel approach to elementary scattering
theory, Nuovo Cimento \textbf{27}, 261 (1963).

\bibitem{Dashen1963} R.F. Dashen, Some extensions of the born approximation for
phase shifts, Nuovo Cimento \textbf{28}, 229 (1963).

\bibitem{Wiringa1995} R.B. Wiringa, V.G.J. Stoks, R. Schiavilla, Accurate
nucleon-nucleon potential with charge-independence breaking, Phys.
Rev. C \textbf{51}, 38 (1995).

\bibitem{Machleidt2001} R. Machleidt, High-precision, charge-dependent Bonn
nucleon-nucleon potential, Phys. Rev. C \textbf{63}, 024001
(2001).

\bibitem{Epelbaum2015} E. Epelbaum, H. Krebs, U.-G. Meissner, Improved chiral
nucleon-nucleon potential up to next-to-next-to-next-to-leading
order, Eur. Phys. J. A \textbf{51}, 53 (2015).

\bibitem{Perez2013} R.N. Perez, J.E. Amaro, E.R. Arriola, Partial-wave analysis of
nucleon-nucleon scattering below the pion-production threshold,
Phys. Rev. C \textbf{88}, 024002 (2013).

\bibitem{Dolgopolov2017} M.A. Dolgopolov, L.A. Minin, V.A. Rabotkin, Approximation
Properties of the Paris Potential of Nucleon-Nucleon Interaction,
Bull. Russ. Acad. Sci.: Phys. \textbf{81}, 1225 (2017).

\end{thebibliography}
\end{document}